\documentclass[amssymb,prb,aps,showpacs,twocolumn,showpacs]{revtex4}
\usepackage{graphicx}
\begin{document}

\title{\textbf{Quantum model of a solid-state spin qubit: Ni cluster on a silicon surface by the generalized spin Hamiltonian and X-ray absorption spectroscopy investigations}}
\author{Oleg V.Farberovich$^{1,2,3}$, Victoria L.Mazalova$^{2}$, Alexander V.Soldatov$^{2}$}
\affiliation{$^1$School of Physics and Astronomy, Beverly and Raymond Sackler Faculty of Exact Sciences,
Tel Aviv University, Tel Aviv 69978, Israel\\
$^2$Research center for nanoscale structure of matter, Southern Federal University,
Zorge 5, 344090, Rostov-on-Don,Russia\\
$^3$Voronezh State University, Voronezh 394000, Russia}
\date{\today}
\begin{abstract}
We present here the quantum model of a Ni solid-state electron spin qubit on a silicon surface with the use of a
density-functional scheme for calculation of the exchange integrals in  the non-collinear spin configurations in the generalized spin Hamiltonian (GSH) with the anisotropic exchange couplings parameters linking the nickel ions with a silicon substrate.
In this model the interaction of a spin qubit with substrate
is considered in GSH at the calculation of exchange integrals $J_{ij}$
of a nanosystem Ni$_7$-Si in the one-electron approach taking
into account chemical bonds of all Si-atoms of a
substrate (environment) with atoms of a Ni$_7$-cluster.
The energy pattern was found from the effective GSH Hamiltonian acting  in the restricted spin space of the Ni ions by the application of the irreducible tensor operators (ITO) technique. In this
article we offer the model of the quantum solid-state N-spin qubit based on the studying of the spin structure and the spin-dynamics
simulations of the 3d-metal Ni clusters on a silicon surface.
The solution of the problem of the entanglement between a spin
states in a N-spin systems is becoming more interesting when considering clusters or molecules with a spectral gap in their
density of states. For quantifying the distribution of the entanglement
between the individual spin eigenvalues (modes) in the spin
structure of the N-spin system we use the density of entanglement(DOE).
In this study we have developed and used the advanced high-precision numerical techniques to accurately assess the details of the decoherence process governing the dynamics of a N-spin qubits interacting with a silicon surface.
We have studied the Rabi oscillations to evaluate the N-spin qubits system
as a function of the time and the magnetic field. We have observed the stabilized
Rabi oscillations and have stabilized the quantum dynamical
qubit state and Rabi driving after a fixed time (0.327 $\mu s$).
The comparison of the energy pattern with the anisotropic
exchange models conventionally used for the analysis of this system
and, with the results of the experimental XANES spectra, shows
that our complex investigations provide a good description of the
pattern of the spin levels and the spin structures of the
nanomagnetic Ni$_7$ qubit. The results are discussed in the view
of the general problem of the solid-state spin qubits and the spin
structure of the Ni cluster.
\end{abstract}
\maketitle                                                                                                                      \section{Introduction}
The promise of quantum computers to solve classically non-computable problems
\cite{i1} has generated the great                                                                                           excitement and much research activity in different areas of physics, mathematics and engineering.
Various physical systems have been proposed for implementation of quantum bits (qubits) in quantum
information processing devices: trapped ions, atoms in QED cavities, magnetic molecules, etc. Among
many candidates, spin-based solid-state systems, such as quantum dots \cite{i2,i21} or spin centers
in host crystals (phosphorus donors in silicon \cite{i3}, NV centers in diamonds \cite{i4,i41}),
constitute attractive candidates for qubits:
these systems are well scalable, can be fabricated and operated by the methods of modern
microelectronics, and advanced spin-resonance techniques are well-suited for the efficient quantum state
manipulation. Thus, it is not surprising that a large number of leading research groups, both theoretical
and experimental, focus their studies on developing and investigating solid-state spin-based qubits.

Since the quantum computing technology has being improved and quantum computers with the nontrivial number of N qubits appear feasible in the near future, an application
of the quantum computers with N$\geq$2000 has ripened \cite{i5}.
The information technologies provide very interesting challenges and an
extremely wide playground in which scientists working in materials science, chemistry,
physics and nano-fabrication technologies
may find stimuli for novel ideas. Curiously, the nanometre scale is the cluster scale.
So we may wonder whether, how or simply
which functional magnetic clusters can be regarded in some ways as the possible components of the
solid-state electron spin qubit, which is the fundamental concept of the quantum
computation. Key challenges in building a quantum computer from spin qubits in physical systems are preparation of arbitrary spin states, implementation of the arbitrary qubit evolution, reading out the qubit states, overcoming of  the decoherence and doing all this on a large scale; that is, with a large number of qubits or a spin system of N-qubits (SSNQ) with the definite spin structure.
For the purposes of implementing quantum computation,
the physical system can be treated as a SSNQ in
which the couplings between the qubits can be controlled externally.
The concept of a SSNQ can be related to the problem of the quantum
spin structure, where the nontrivial applications may exist for computers with a limited
number N of a qubits. The precise relationship between the type of the entanglement
and the distribution of the coupling strengths
in the SSNQ can be strongly dependent
on external parameters, such as applied magnetic fields and temperature.
In this context, systematic studies of the relationship
between the amount and nature of the entanglement and the spin structure of the SSNQ
has been pursued in order to identify the optimal
spin structures to create specific types of the entanglements \cite{i6}.

Spins are alternative complementary to charges as degrees of freedom to encode information.
Recent examples,
like for instance the discovery and application of Giant magnetoresistance in Spintronics,
have demonstrated the
efficient use of a spins for information technologies \cite{i8}.
Moreover, spins are intrinsically quantum entities and they have
therefore been widely investigated in the field of the quantum-information processing.
The cluster nanomagnets of a transition metal are
real examples of finite spin-clusters (0D), and therefore they constitute a new
benchmark for testing models of the interacting
quantum objects. New physics of the cluster magnets feeds hopes of certain prospective applications,
and such hopes pose the problem
of understanding, improving, or predicting desirable characteristics of these materials.
The magnetic transition metal
nanostructures on a non-magnetic substrates have attracted recently large attention
due to their novel and the unusual magnetic
properties \cite{i9}. The supported clusters experience both the reduction of the local coordination number, as in free
clusters, as well as the interactions with the electronic degrees of freedom of the substrate, as in embedded clusters. The
complex magnetic behavior is usually associated with the competition of the several
interactions, such as interatomic exchange and
bonding interactions, and in some cases the non-collinear effects, which can give rise to
the several metastable states close in energy.
The ground state can therefore be easily tuned by external action giving rise to the
switching between different states \cite{i10}.

Therefore the goal is ambitious: it is not just a matter to store information in a $3d$-metal
cluster on a non-magnetic
substrate, but we may think to process information with a cluster and then to
communicate information at the clusters containg
from magnetic $3d$-metal atoms on a silicon surface.
Among the various candidates for a solid-state qubits, spins
have been of the particular interest due to their relative robustness
to decoherence compared with other degrees of freedom, such as
a charge. The most coherent solid-state systems investigated so far
are the spins of well isolated donors in bulk 28Si, which
produce coherence times (T2) of up to seconds (extrapolated)
for the electron spin and minutes for the nuclear spin, which are
comparable to those of a ion trap qubits \cite{i7}. The problem of decoherence
comprises our main motivation to study the decoherence
dynamics of a N--qubits system \cite{i11,i12,i13}.
From the experimental point of view, the coherent transition from a coherent to
an incoherent dynamics can be probed by the observation of Rabi oscillations
between the quantum states of the a spin precessing
in a static magnetic field \cite{i161}. A related problem in the context of the present study is the Rabi oscillations in a
SSNQ.  Here we suggest for construction of the spin qubit the Ni small clusters on a silicon surface. The stabilization of Rabi
oscillations and the maximum of the entanglement were discussed in this SSNQ.
In recent years, the entanglement has attracted the attention of a many physicists working in the area of a quantum
mechanics \cite{i14,i15,i16,i161}. This is due to the ongoing research in the area of quantum information \cite{i17}. Theoretical
studies are also important in the context of spin interactions inside structured reservoirs such as a metal cluster on a nonmagnetic
surface. Ni is the unique among the transition-metal adatoms, because
its half-filled valence configuration ($3d^84s^2$) yields
strong interatomic bonding leading to magnetic frustration.
We apply our method to the Ni clusters deposited on a Si(111) surface.
In the present work we study the entanglement
between the spin states in the spin spectrum.
In our model, a spin state interacts with the spin structure of the continuum at the temperature interval 0 -- 300 K, and the entanglement properties between the spin states in the spin structure are considered.
Using the global entanglement as a measure of the entanglement, we
derive a pair of distributions that can be interpreted as the density of
entanglement in the terms of all the eigenvalue of the spin
spectrum. This distribution can be calculated in terms of the spectrum
of the spin excitation of the Ni${_7}$-cluster on a
Si (111)-surface.
With these new measures of the entanglement we can study in detail
the entanglement between the spin modes in the spin structure.
The method developed here, in terms of the entanglement distributions, can
also be used when considering various types of the structured reservoirs \cite{i18}.
The low-lying excited states of a magnetic system are generally described in the terms
of a general spin-Hamiltonian \cite{i18,ITO20}.
For a magnetic system with the many spin sites, this phenomenological Hamiltonian
is written as a sum of a pair-wise spin exchange
interactions between adjacent spin sites in a cluster and a surface.

As the most properties depend on the unique features of the local atomic structure,
the diagnostic methods to the control structural
parameters, such as the distances between the atoms, with the very high precision
are required. In this work X-Ray Absorption
Near-Edge Structure (XANES) spectroscopy was used both to study the adsorption
geometry and to get information concerning the
electronic properties of the deposited Ni$_7$ clusters \cite{XANES1,XANES2}.
XANES spectroscopy is now a powerful tool of
investigation of the atomic and electronic structure of different classes of
materials in condensed state. The XANES spectroscopy
has essential advantages as compared with other methods of atomic structure analysis.
For example, in contrast to X-ray and
neutron crystallography, XANES spectroscopy can be used for investigation of materials
without long order in atoms arrangement.
Extended X-ray Absorption Fine Structure (EXAFS) spectroscopy also allows studying
the compounds without long order, but it can
give only information about the coordination numbers and interatomic distances,
while XANES is very sensitive to the both small
bond distances and bond angles variations. For example, XANES allows determination
of the interatomic distances with 0.02
Angstrom accuracy as well as the study of bonding angles with the accuracy up to
several degrees. Thus, on the basis of XANES
analysis it is possible to determine the full 3D atomic structure of studied materials.
Previously, we have used the method of X-ray absorption spectroscopy for a comprehensive
study of copper nanoclusters and diluted magnetic semiconductors \cite{XANES1,XANES2}.
The knowledge of the properties of nanoclusters made it possible to obtain information
on how the transition from the atom or cluster to the solid state can occur.
In order to extract the necessary information from the experimental XANES spectra
one needs to perform theoretical analysis. In
the present research the theoretical XANES spectra was simulated using FEFF9.0
program code \cite{XANES3,XANES4}. FEFF9.0 code is
based on the real space full-multiple scattering theory. The code uses the cluster
approach for XANES spectra calculations and,
therefore (in contrast to program based on the band structure calculations),
can be applied for study of compounds without long
order in atoms distributions, like clusters. The electronic structure of investigated
clusters was analyzed on the basis of the
density functional theory (DFT) implemented in the ADF2013 code \cite{XANES5,XANES6}.
\section{THE THEORETICAL APPROACH}                                                                                                                                                          We are using here a quantum algorithm with the three distinct steps: the calculation of the magnetic
properties for the qubit; encoding a spin
wavefunction into the qubits (the spin structure); a spin-dynamics  simulating its time evolution.                                    In order to give a theoretical description of a magnetic cluster we exploit the irreducible
tensor operator technique \cite{ITO20}.                                                                                      Let us consider a spin cluster of an arbitrary topology formed from an arbitrary number of
the magnetic sites, $N$, with a local spins S$_1$, S$_2$,..., S$_N$ which, in general, can
have a different values. A successive spin coupling scheme is adopted:
\begin{equation}                                                                                                                 S_{1} + S_{2} = S^{(2)}, S^{(2)} + S_3 = S^{(3)},..., S^{(N-1)} + S_N = S,                                                       \end{equation}                                                                                                                   where $S$ represents the complete set of a intermediate spin quantum numbers $S^{(k)}$, with $k$=1,2,...,N-1.
The eigenstates
$\mid\it SM\rangle$ of the spin-Hamiltonian are given by the linear combinations of the basis states $\mid                           S^{(\mu)}M^{(\mu)}\rangle$:                                                                                                      \begin{equation}\label{spinvf}                                                                                                                 \mid\it SM\rangle = \sum_{\mu=1}^N\langle c_{\mu}\mid\it SM\rangle\mid S^{(\mu)}M^{(\mu)}\rangle,                                \end{equation}                                                                                                                   where $M=-S,...,S$ and the coefficients $\langle c_{\mu}\mid\it SM\rangle$ can be evaluated once
the spin-Hamiltonian of the system has been diagonalized. Each term of the spin-Hamiltonian
can be rewritten as a combination of the irreducible tensor
operators technique \cite{ITO20}. The work \cite{GSH19} focuses on the main physical interactions which determine the spin-Hamiltonian and to rewrite them in the terms of the ITO's. The exchange part of the
spin-Hamiltonian is introduced:
\begin{equation}\label{GSH}                                                                                                                 \widehat H_{spin} = \widehat H_0 + \widehat H_{BQ} + \widehat H_{AS} + \widehat H_{AN}.                                          \end{equation}                                                                                                                   The first term $\hat H_0$ is the Heisenberg Hamiltonian, which represents the isotropic
exchange interaction,
$\widehat H_{BQ}$ is the biquadratic exchange Hamiltonian, $\widehat H_{AS}$ is the
antisymmetric exchange Hamiltonian, and
$\widehat H_{AN}$ represents the anisotropic exchange interaction.
Conventionally, they can be expressed as follows \cite{GSH19}:
\begin{equation}\label{eq4}                                                                                                                 \widehat H_0 = -2\sum_{i,f}J_{if}\widehat{S}_i\widehat{S}_f                                                                      \end{equation}                                                                                                                   \begin{equation}                                                                                                                 \widehat H_{BQ} = -\sum_{i,f}j_{if}(\widehat{S}_i\widehat{S}_f)^2                                                                \end{equation}
\begin{equation}                                                                                                                 \widehat H_{AS} = \sum_{i,f}{\bf G}_{if}[\widehat{S}_i\times \widehat{S}_f]                                                      \end{equation}
\begin{equation}                                                                                                                 \widehat H_{AN} = -2\sum_{i,f}\sum_{\alpha}J_{if}^{\alpha}\widehat{S}_i^\alpha\widehat{S}_f^\alpha                               \end{equation}                                                                                                                   with $\alpha$ = x, y, z.

We can add to the                                                                                                                exchange Hamiltonian the term due to the axial single-ion anisotropy:
\begin{equation}                                                                                                                 \widehat H_{ZF} =\sum_{i}D_i \widehat{S}^2_z(i)                                                                                  \end{equation}                                                                                                                   where $J_{if}$ and $J_{if}^{\alpha}$ are the parameters of the isotropic and anisotropic
exchange iterations, $j_{if}$ are the
coefficients of the biquadratic exchange iterations, and ${\bf G}_{if}$=-${\bf G}_{fi}$
is the vector of the antisymmetric
exchange. The terms of the spin-Hamiltonian above can be written in the terms of the ITO's.

In this paper, we use the results of a first-principles calculations of the exchange
parameters $J_{ij}$.
Here, we introduce  the scenario \cite{MasFar21} to construct the parameters within the classical
spin model that contains the interactions,
in principle, up to an arbitrary order. Our development of a non-collinear method is
based on a semi-relativistic
first-principle calculations of the energy in the framework of the density functional
theory (DFT) within the linear
combination of the atomic orbitals (LCAO) method \cite{Kub22}.
We suggested a straightforward approach that allows the direction of the magnetic
moment of the any atom to be fixed
by using only a on-site information \cite{Mas23}. Thus, we can obtain a sufficiently large
number of a states with the different
non-collinear magnetic ordering and map them onto an effective spin model by
using least-squares methods.

The Heisenberg and biquadratic exchange are the isotropic interactions.
In fact, the corresponding Hamiltonians can be
described by the rank-0 tensor operators and thus these have the non zero matrix elements
only with the states with the same total spin quantum
number $S$ ($\Delta S$,$\Delta M$=0). The representative matrix can be decomposed
into the blocks depending only on the value of $S$
and $M$. The all anisotropic terms are described by the rank-2 tensor operators which have
the non zero matrix elements between state with
$\Delta S$=0,$\pm 1$,$\pm 2$ and their matrices can not be decomposed into the blocks
depending only on the total spin $S$ in account of
the $S$--mixing between spin states with different $S$. The single-ion anisotropy
can be written in the terms of the rank-2 single site
ITO's \cite{GSH19}. Finally, the antisymmetric exchange term is the sum of
the ITO's of the rank-1.

The ITO technique has been used to design the MAGPACK software \cite{ITO20},
a package to calculate the energy levels,
bulk magnetic properties, and inelastic neutron scattering spectra of the high nuclearity
spin clusters that allows studying
efficiently properties of a nanoscopic magnets.
\subsection{Calculation of the spin structure of the N-spin system}                                                                  One of the major challenges in quantum computing is to identify a system that can be scaled
up to the number of the qubits
needed to execute the nontrivial quantum algorithms. Peter Shor’s algorithm \cite{i5} for finding
the prime factors of the numbers
used in the public-encryption systems (numbers that typically consist of more than a hundred digits)
would likely
require a quantum computer with the several thousand qubits. Depending on the error correction
scheme appropriate to
the particular computer, the required number could be much larger.
The solid-state spin quantum computers
may be more likely candidates. In this work we offer the model of the quantum computer
based on studying of the spin structure of the
 3d-metal Ni clusters on a silicon surface.
In our model the total electronic structure can be written as a sum of a
non-spin-polarized charge DOS $n_{DOS}(\epsilon)$, and the spin DOS, $S_{DOS}(\epsilon)$,                                                                          \begin{equation}\label{DOS}                                                                                                                                                n(\epsilon) = n_{DOS}(\epsilon) + S_{DOS}(\epsilon)                                                                                                             \end{equation}                                                                                                                                                  The first part $n_{DOS}(\epsilon)$ is connected with a one-electron structure and
can be received by any DFT method.
The second part $S_{DOS}(\epsilon)$ is purely a spin structure and is defined
only proceeding from the spin model of a cluster.
For division in an experimental spectra of the deposits from a purely spin
states and from the excitations connected with the one-electron
transitions, is reasonable within uniform approach to the receive electronic structure
of a cluster and then, using the one-electron
data for calculation of the exchange integrals, to
calculate a spin structure by the ITO method within the generalized spin
Hamiltonian $\widehat H_{spin}$.
We have developed the idea of spins as a degree of freedom, with which models are built \cite{LDiV24}.
The spin magnetic moment
due to the exchange interaction is                                                                                                                               \begin{equation}\label{Ms}                                                                                                                                                 {\bf M}_s=-2\langle \hat {\bf S}\rangle \mu_B/\hbar,                                                                                                             \end{equation}                                                                                                                                                   where                                                                                                                                                            $$                                                                                                                                                               \langle \hat S\rangle = \sqrt{\langle \hat S_x\rangle^2+\langle \hat S_y\rangle^2+\langle \hat S_z\rangle^2}                                                     $$                                                                                                                                                               is the spin structure. With the spin-Hamiltonian $\widehat H_{spin}$ result (\ref{GSH}) we can obtain
the quantum mechanical expectation values:
\begin{equation}                                                                                                                                                 \langle \hat S_x\rangle=\langle SM\mid\hat S_x\mid SM\rangle,                                                                                                    \end{equation}
\begin{equation}                                                                                                                                                 \langle \hat S_y\rangle=\langle SM\mid\hat S_y\mid SM\rangle,                                                                                                    \end{equation}
\begin{equation}                                                                                                                                                 \langle \hat S_z\rangle=\langle SM\mid\hat S_z\mid SM\rangle.                                                                                                    \end{equation}                                                                                                                                                   With the algebra of the spin operators we can obtain the expectation values for $\widehat H_{spin}$:                                                                              \begin{equation}                                                                                                                                                 \langle \hat S_x\rangle=\frac{1}{2}\sum_{\mu=1}^{N}c_\mu^2A_{\mu},                                                                                               \end{equation}
\begin{equation}                                                                                                                                                 \langle \hat S_y\rangle=-\frac{i}{2}\sum_{\mu=1}^{N}c_\mu^2B_{\mu},                                                                                              \end{equation}
\begin{equation}                                                                                                                                                 \langle \hat S_z\rangle=\sum_{\mu=1}^{N}c_\mu^2M^{(\mu})                                                                                                         \end{equation}                                                                                                                                                   where                                                                                                                                                            $$                                                                                                                                                               A_{\mu}=\sqrt{S^{(\mu)}(S^{(\mu)}+1)-M^{(\mu)}(M^{(\mu)}+1)}+                                                                                                    $$                                                                                                                                                               $$                                                                                                                                                               +\sqrt{S^{(\mu)}(S^{(\mu)}+1)-M^{(\mu)}(M^{(\mu)}-1)}                                                                                                            $$                                                                                                                                                               and                                                                                                                                                              $$                                                                                                                                                               B_{\mu}=\sqrt{S^{(\mu)}(S^{(\mu)}+1)-M^{(\mu)}(M^{(\mu)}+1)}-                                                                                                    $$                                                                                                                                                               $$                                                                                                                                                               -\sqrt{S^{(\mu)}(S^{(\mu)}+1)-M^{(\mu)}(M^{(\mu)}-1)}                                                                                                            $$
The spin DOS $S(\epsilon)$ can be done by inserting the following functions:                                                                                     \begin{equation}                                                                                                                                                 S_x(\epsilon)=\sum_{\mu=1}^{N}c_\mu^2A_{\mu}\delta(\epsilon-\epsilon_\mu),                                                                                       \end{equation}

\begin{equation}                                                                                                                                                 S_y(\epsilon)=\sum_{\mu=1}^{N}c_\mu^2B_{\mu}\delta(\epsilon-\epsilon_\mu),                                                                                       \end{equation}
\begin{equation}                                                                                                                                                 S_z(\epsilon)=\sum_{\mu=1}^{N}c_\mu^2M^{(\mu)}\delta(\epsilon-\epsilon_\mu).                                                                                     \end{equation}                                                                                                                                                   The spin DOS is                                                                                                                                                  \begin{equation}                                                                                                                                                 S_{DOS}(\epsilon) = \sqrt{S_x^2(\epsilon)+S_y^2(\epsilon)+S_z^2(\epsilon)}                                                                                       \end{equation}
\subsection{Calculation of the magnetic properties}
Once we have the energy levels, we can evaluate a different thermodynamic properties
of the system as the magnetization,
the magnetic susceptibility, and the magnetic specific heat.
Since in  further researches the anisotropic part of GSH will be only scalar, the magnetic properties                                  of the anisotropic system do not depend on the direction of the magnetic field.
Thus we can consider the external magnetic
field $H_z$ directed along arbitrary axis $\it z$ of the cluster coordinate frame that is
chosen as a spin quantization axis. In
this case the energies of the system will be
$\epsilon_{\mu}(M_s) +   g_e\beta M_s H_z$, where $\epsilon_{\mu}(M_s)$
are the eigenvalues of the GSH containing the magnetic exchange and
the double exchange contributions
(index $\mu$ runs over the energy levels with given total spin protection ${\bf M}_s$).                                                                          Then the partition function in the presence of the external magnetic field is given by:                                                                          \begin{equation}\label{Z}                                                                                                                                                 Z(H_z) = \sum_{M_s,\mu}\exp[-\epsilon_\mu(M_s)/kT]\sum_{M_s}\exp[-g_e\beta M_s H_z/kT]                                                                           \end{equation}                                                                                                                                                   Using this expression one can evaluate the magnetic susceptibility $\chi$ and the magnetization
$M_s$ by the standart thermodinamic definitions:                                                                                                                                                     \begin{equation}\label{chi}                                                                                                                                                 \chi = \left(\frac{\partial {\bf M}_s}{\partial {\bf H}}\right)_{{\bf H}\to 0}                                                                                   \end{equation}
\begin{equation}\label{M}                                                                                                                                                 {\bf M}_s({\bf H})=NkT\frac{\partial{\ln Z}}{\partial \bf H}                                                                                                     \end{equation}
\subsection{The spin-dynamics simulations}
Broadly speaking, a quantum computer is a physical system that can be initialized
to some known state $\mid\Psi(t_i)\rangle =
\mid\Psi(t_0)\rangle$, and whose dynamics can be controlled so as to induce
any unitary transformation of the state vector
$\mid\Psi(t_i)\rangle =\hat U\mid\Psi(t_0)\rangle$. In the standard approach,
the computational process is given by the unitary
time-evolution operator $\hat U$ of the state vector, and is driven by the
application of the external stimuli. Under the assumption
of a small time step $\Delta t$ we can expand the time evolution operator                                                                                        $$                                                                                                                                                               \hat U(t+\Delta t,t)\approx(I-iH_{spin}|\Delta t/\hbar)+O(\Delta t^2)                                                                                            $$                                                                                                                                                                Probably the best starting point for a analytical considerations of the dynamics of the quantum
spins is the Heisenberg equation of a motion \cite{W25}                                                                                                                                         \begin{equation}                                                                                                                                                 i\hbar\frac{\partial\langle\widehat {\bf S}\rangle}{\partial t}=[\langle\widehat{\bf S}\rangle,\widehat                                                          H_{spin}]+i\alpha[\langle\widehat{\bf S}\rangle,[\langle\widehat {\bf S}\rangle,\widehat H_{spin}]],                                                             \end{equation}                                                                                                                                                   where $\alpha$ is the damping parameter ($\alpha\geq0$).
In \cite{W26} was demonstrated that                                                                                                                             \begin{equation}                                                                                                                                                 \frac{i}{\hbar}[\langle\widehat {\bf S}\rangle,\widehat H_{spin}]=\langle\widehat {\bf S}\rangle\times\frac{\partial \widehat                                    H_{spin}}{\partial\langle\widehat {\bf S}\rangle}+O(\hbar)                                                                                                       \end{equation}                                                                                                                                                   and thus                                                                                                                                                         \begin{equation}                                                                                                                                                 \frac{\partial\langle\widehat {\bf S}\rangle}{\partial t}=-\langle\widehat {\bf S}\rangle\times\frac{\partial \widehat                                           H_{spin}}{\partial\langle\widehat {\bf S}\rangle}+\alpha(\langle\widehat {\bf S}\rangle\times\frac{\partial\langle\widehat {\bf                                  S}\rangle}{\partial t})                                                                                                                                          \end{equation}                                                                                                                                                   This time-dependent equation is similar to Landau-Lifshitz-Gilbert (LLG) equation \cite{W27}
for the spin magnetic moment ${\bf M}_s$ (\ref{Ms}) dynamics.                                                                                                                                       \begin{equation}                                                                                                                                                 \frac{\partial{\bf M}_s}{\partial t}=-2{\bf M}_s\times\mathfrak{H}_{eff}+                                                                                        \alpha{\bf M}_s\times\frac{\partial{\bf M}_s}{\partial t}                                                                                                        \end{equation}                                                                                                                                                   In order to solve this equation, as an effective magnetic field, we take the variational derivative of the energy with respect to the magnetization.
$$                                                                                                                                                               \mathfrak{H}_{eff} =-\frac{\delta \mathfrak{F}}{\delta{\bf M}_s}                                                                                                 $$                                                                                                                                                               where $\mathfrak{F}$ is the free energy of the magnetic nanosystem                                                                                               $$                                                                                                                                                               \mathfrak{F}=-Nk_BT\ln Z(H_z)                                                                                                                                    $$                                                                                                                                                               with the partition function (\ref{Z}). Here                                                                                                                           $$                                                                                                                                                               {\bf M}_s = -\frac{\partial \mathfrak{F}}{\partial{\bf H}}                                                                                                       $$                                                                                                                                                               and                                                                                                                                                              $$                                                                                                                                                               \chi=\frac{\partial{\bf M}_s}{\partial{\bf H}},                                                                                                                  $$                                                                                                                                                               where ${\bf H}$ is the magnetic field.                                                                                                                               Thus                                                                                                                                                             $$                                                                                                                                                               \mathfrak{H}_{eff} =-\frac{{\bf M}_s^{(0)}}{\chi^{(0)}}                                                                                                          $$                                                                                                                                                               calculated by formulas (\ref{chi}) and (\ref{M}).                                                                                                                            We have derived a general form of the time-dependend spin equation for a system of the spins                                                                             precessing in an effective magnetic field with specifying the all interactions in GSH (\ref{GSH}).
\subsection{The entanglement in the N-spin system}
Identifying and measuring of the entanglement in multi-spin systems presents various complications.
Apart from the case of a two-qubit system, where the entanglement can be identified both for a pure and a
mixed state \cite{Woo28,Uhl29,Aud30,Woo31}, a multi-qubit entanglement is
an open problem \cite{VC32,WN33,CH34}.
For the analysis that follows, we will be using the measure of the recently
proposed density of entanglement.
Based on the residual entanglement \cite{CH34}, we present the global
entanglement of a N-spin state
as collective measures of a multi-particle entanglement.
This measures were introduced by Meyer and Wallach \cite{MW35}.
The Meyer–Wallach (MW) measure is written in the Brennen form \cite{Bren36}                                                                                              and we use the Q-measure \cite{MW35}, which corresponds, for a cluster with
N qubits, to the average purity
of the reduced density matrices of each qubit:                                                                                                                    \begin{equation}                                                                                                                                                 Q(\psi)=\sum\limits_{k=1}^{N}2[1-Tr(\rho_k^2)]                                                                                                                   \end{equation}                                                                                                                                                   where $\rho_k$ is the reduced density matrix for $k$-th qubit.                                                                                                   The problem of the entanglement between spin states in the N-spin systems
become more interesting when the clusters with a spectral gap in their density
of the states are considered.
For quantifying the distribution of
the entanglement between the individual spin eigenvalues in the spin structure of
the N-spin system we use the density of the entanglement
(DOE) \cite{LG37}. The density of the entanglement
$\varepsilon(\epsilon_{\mu},\epsilon_{\lambda},\epsilon)d\epsilon$
gives the entanglement between the spin eigenvalue $\epsilon_{\mu}$ and
the spin eigenvalue $\epsilon_{\lambda}$ in the energy
interval $\epsilon_{\mu}$ to $\epsilon_{\mu}+d\epsilon_{\mu}$.
One can show that the entanglement distribution can be written in the terms of
a spectrum of a spin excitation
\begin{equation}                                                                                                                                                 S(\epsilon_{\lambda},\epsilon)=                                                                                                                                  \left|c_\lambda\right|^2\delta(\epsilon-\epsilon_\lambda)                                                                                                        \end{equation}                                                                                                                                                   and                                                                                                                                                              \begin{equation}                                                                                                                                                 \varepsilon(\epsilon_{\mu},\epsilon_{\lambda},\epsilon)=                                                                                                         2S(\epsilon_{\lambda},\epsilon)S(\epsilon_{\mu},\epsilon)                                                                                                        \end{equation}                                                                                                                                                   where the coefficient $\langle c_{\lambda}\mid\it SM\rangle$ is the eigenvector (\ref{spinvf}) of
the spin-Hamiltonian (\ref{GSH}) of the cluster.
Thus, the entanglement distributions can be derived from the excitation spin spectrum                                                                                \begin{equation}                                                                                                                                                 Q(\epsilon)=1-\frac{2\Delta^2}{\pi^2N}\sum\limits_{\mu=1}^{N}                                                                                                    \frac{\left|c_\mu\right|^2}{(\epsilon-\epsilon_{\mu})^2+\Delta^2}                                                                                                \sum\limits_{\lambda=\mu+1}^{N}\frac{\left|c_\lambda\right|^2}{(\epsilon-\epsilon_{\lambda})^2+\Delta^2}                                                         \end{equation}                                                                                                                                                   where $\Delta$ is the Lorentzians width.
Though the very nature of the entanglement is the purely quantum mechanical, we saw that it can persist for the macroscopic systems and survive even in the thermodynamical limit. In this section we discuss how it behaves at the finite
temperature of the thermal entanglement. The states in the N-spin system, describing a
system in the thermal equilibrium states, are
determined by the generalized spin-Hamiltonian and the thermal density matrix                                                                                        \begin{equation}                                                                                                                                                 \rho(T)=\frac{\exp(-H_{spin}/kT)}{Z(H_z)}                                                                                                                        \end{equation}                                                                                                                                                   where $Z(H_z)$ is the partition function of the the N-spin system.                                                                                                   The thermal entanglement is                                                                                                                                      \begin{equation}                                                                                                                                                 Q(\epsilon,T,H_z)=1-\frac{2\Delta^2}{\pi^2NZ(H_z)^2}\sum\limits_{\mu=1}^{N}                                                                                      \frac{\left|c_\mu\right|^2\exp[-\epsilon_\mu/kT]}{(\epsilon-\epsilon_{\mu})^2+\Delta^2}\times                                                                    \end{equation}                                                                                                                                                   $$                                                                                                                                                               \sum\limits_{\lambda=\mu+1}^{N}\frac{\left|c_\lambda\right|^2\exp[-\epsilon_\lambda/kT]}{(\epsilon-\epsilon_{\lambda})^2+\Delta^2}                               $$
\subsection{Calculation of the density of states $n_{DOS}(\epsilon)$}
The total density of states $n_{DOS}(\epsilon)$ of Ni$_7$ clusters
was calculated by the all-electron density functional theory
(DFT) approach implemented in the ADF2013 code \cite{XANES5,XANES6}.                                                                                                 The main point of the density functional theory is that the potential acting on each electron of all the other electrons in the molecule or crystal, depends only on the electron density of the ground state and its gradient. Thus, we can apply the one-electron formulation of a system of N interacting electrons by introducing the appropriate local exchange-correlation potential $ V_{XC}(r)$, in addition to any external potentials $ V_{ext} (r) $ and the Coulomb potential of the electron cloud $ V_C(r) $ and it has the form:
\begin{equation}
(-\frac{1}{2}\nabla^2+V_{ext}(\vec r)+V_c(\vec{r})+V_{xc}(\vec{r}))\phi_i(\vec{r})=\varepsilon_i\phi_i(\vec{r})
\end{equation}
The one-electron molecular orbitals $ \phi_i $ with the appropriate orbital energies $ \varepsilon_i $ define precise electronic charge density and, in principle, give access to all the properties, because they are all expressed in terms of density functional, especially energy. In addition, they allow us to represent the system as a set of independent electron orbitals.						
In our calculations the electronic configuration of the $ Ni_7 $ cluster was described by the polarized
triple zeta (TZP) basis set of the Slater-type orbitals.
The total DOS was obtained using the BLYP exchange functional which is
equivalent to Becke (exchange)\cite{XANES7} and
Lee-Yang-Parr (LYP correlation)\cite{XANES8}.
The calculations were done for the high-spin state of Ni$_7$ cluster.

\section{Results and discussion}
Here we present the results of the application of our theoretical and
experimental approach to study of the N-qubits from
seven nickel atoms interacting with a Si
(111)-surface in the terms of numerically solvable DFT-models for the exchange
integrals $J_{ij}$ and we exploit
the ITO technique \cite{GSH19} for calculation of
the spin structure of the N-qubits system. In
practice, this model was applied to the nanosystem (Ni$_7$-cluster on Si (111)-surface)
and provides an understanding of the
chemical bonding of the nickel nanoparticles with the silicon substrate.
It is advantageous to have access to sufficiently accurate DFT-total
energies to compare the different competing spin
configurations. In the most first-principles calculations of the magnetic
systems only ferromagnetic or some antiferromagnetic
states were considered. Along with these collinear magnetic
configurations, many nanomagnetic clusters show the non-collinear ground
states, such as conical or spin spirals or commensurate superpositions
of several spiral spin-density waves. To access such
states from the first principles, the vector-spin DFT,
which treats the magnetization density as a vector field (and
not as a scalar field as in the collinear DFT calculations) has to be applied \cite{Kub22,Mas23}.
The ab initio magnetic interactions are mapped onto
a certain parametric GSH model (in the simple cases,
the Heisenberg-Dirac-Van Vleck (HDVV) model), which is
studied by using the parameters extracted from the microscopic evaluation \cite{i10}.
Although one of the fundamental parameters
is the magnetic moment of the single cluster, ab initio studies
of its origin are scarce, so little is known about the
cluster spin structure. One of the reasons for the lack of
ab initio studies is the number of the metal ions as spin
sources. The LCAO methods, including the electronic and the spin interactions
such as the vector-spin DFT, are applicable for these clusters
owing to large number of their active electrons and active spaces, and, to the
quantum chemical description of the magnetic anisotropy of the
nanosystems. Our computations of the electronic and the magnetic
properties are based on the LCAO method in the framework of DFT and
have been performed with the package SIESTA \cite{SIESTA}
which was adapted for an investigation of the non-collinear magnetic systems.
Starting from the generalized gradient approximation (GGA)
for the spin-polarized systems, we apply our the on-site constraint method
for the systems with the arbitrary magnetic structures to determine
the ground state and a set of a excited magnetic configurations. To
describe the atoms in the SIESTA code, we generated the
pseudopotentials for the atomic elements according to the Troullier and
Martins procedure \cite{TM38} with the 3p-semicore states for the Ni atoms.
A double-$\zeta$ polarized basis set has been used
for the Si atoms and a triple-$\zeta$ polarized basis set
for the Ni atoms. The exchange-correlation functional
PBE \cite{Per39} was employed. For the real space grid,
we set a uniform mesh corresponding to an energy cut-off of 200 $Ry$.
We wish to determine the exchange interaction parameters from
the first principles by calculating the appropriate total energies and
mapping these results onto the Heisenberg model $\widehat H_{spin}$.
The orientational energy dependence (i.e., the dependence
of the mutual classical spin orientations) can be interpreted in terms of
a classical spin system with an effective spin
Hamiltonian. The magnetic state of N atoms can be characterized by the array                                                    $\{{\bf M}_s^{(i)}\}_{i=1...N}$                                                                                                 where                                                                                                                           ${\bf M}_{s}^{(i)}=\mu_{i}{\bf S}_{i}$                                                                                          is the spin magnetic moment of a particular ion $i$ and $\mu_i$ is
the magnitude of the magnetic moment.
In general, the energy of a spin system up to the second order
with respect to the spin operators (or the classical spin values)
includes the exchange-interaction and the magnetic-anisotropy terms.
Since the spin–orbital interaction \cite{Kort18} is omitted,
the anisotropy terms do not appear in the present calculations and,
therefore, the exchange interaction can be represented
solely by the isotropic HDVV term as shown in Eq.(\ref{eq4}),
where $J_{if}$ is the isotropic exchange coupling constant between
spins $i$ and $f$.                                                                                                              \begin{equation}                                                                                                                E(\{{\bf S}_i\},\{J_{if}\}) = \sum_{i>f}J_{if}{\bf S}_i\cdot{\bf S}_f                                                           \end{equation}                                                                                                                  To obtain the exchange coefficients $J_{if}$, we fit the total
energy $E(\{{\bf S}_i\},\{J_{if}\})$ of the spin system, Eq.(35),
to the orientational potential relief by using the $\chi^2$ method.
For calculation composed $\widehat H_{AS}$ and
$\widehat H_{AN}$ in Eq.(3) we used a technique \cite{Bor42}.

\begin{figure}[tbp!]                                                                                                            \begin{center}                                                                                                                  \includegraphics[height=12cm, width=9cm]{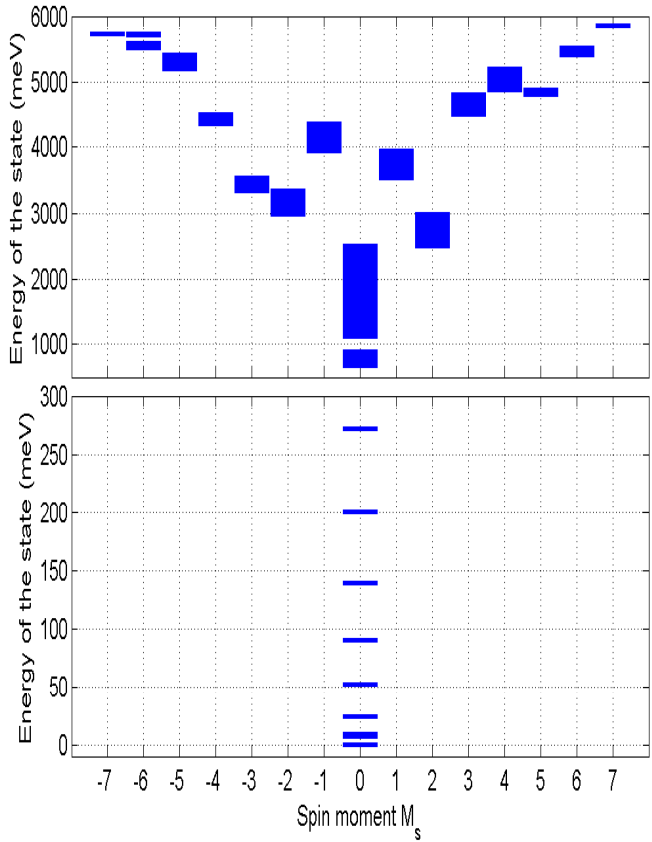}                                                                             \end{center}                                                                                                                    \caption{(Color online) The anisotropic spin-Hamiltonian spectra of Ni$_7$ cluster on a silicon surface.} \label{fig.1}                                                                                \end{figure}

\begin{figure}[tbp!]                                                                                                            \begin{center}                                                                                                                  \includegraphics[height=11cm, width=9.5cm]{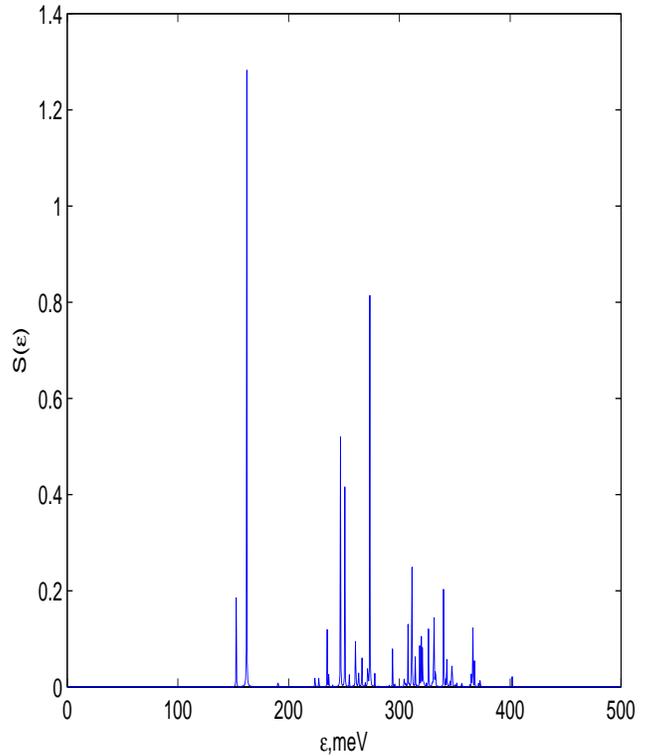}                                                                     \end{center}                                                                                                                    \caption{(Color online) The projected spin DOS $S(\epsilon)$ of Ni$_7$ cluster on a silicon (111)-surface } \label{fig.2}         \end{figure}
\begin{figure}[tbp!]                                                                                                            \begin{center}                                                                                                                  \includegraphics[height=12cm, width=10cm]{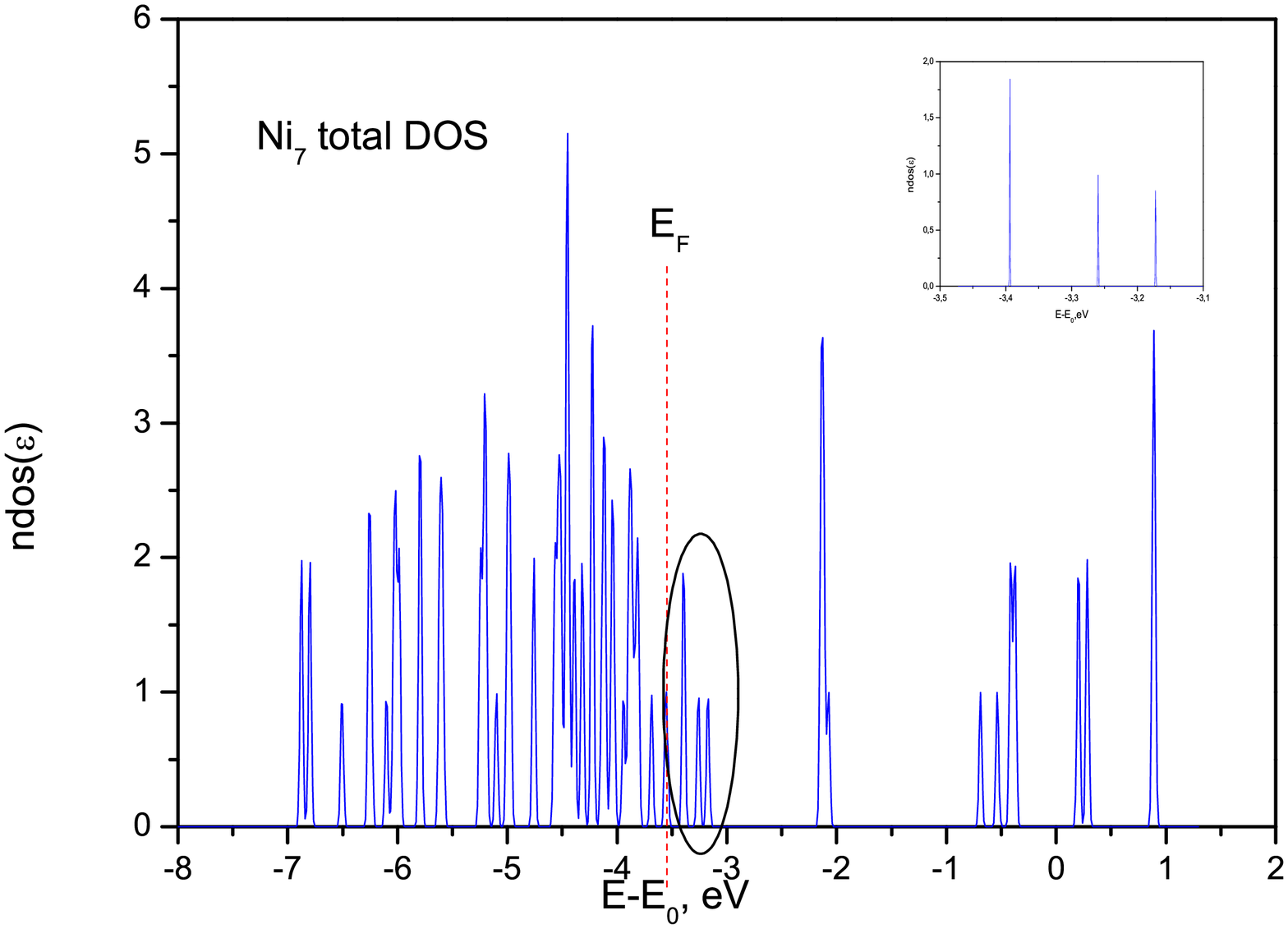}                                                                             \end{center}                                                                                                                    \caption{(Color online) The projected total one-electron $n_{DOS}(\epsilon)$ of Ni$_7$ cluster on a silicon surface. Inset:     expanded view of the first LUMOs.} \label{fig.3}                                                                                \end{figure}

The calculated (within the GSH model) spin levels of Ni$_7$ are shown
in Fig.\ref{fig.1}. They are grouped according to
the spin moment $M_s$. A peculiar feature of the energy
pattern is the presence of the levels belonging to $M_s$ = 0 (the ground
state and the low lying excited states) and $M_s$ = $\pm(1\div7)$
(the highly lying excited states) separated by a small gap $\Delta$,
and the sets of the excited levels ($\geq$ 300 meV) are well separated
from the low-lying levels. The exchange integrals describes the
interaction between the spins of the neighboring atoms. From the analysis of the
calculated values of the exchange integrals $J_{if}$ of
the nickel cluster on a silicon surface it follows that the greatest
values 14.2 - 10.8 meV turn out only for the nearest neighbors
further their values sharply decrease (approximately by 3
orders). In this regard our estimates show that the cluster
from seven atoms well describes the spin structure
of a nickel surface.
In our model the total electronic structure can be written as
a sum of a non-spin-polarized,charge DOS $n_{DOS}(\epsilon)$,
and a spin DOS, $S_{DOS}(\epsilon)$. The spin DOS S($\epsilon$) of Ni$_7$ cluster on a silicon surface
is present on Fig.\ref{fig.2}. The Fig.\ref{fig.3} shows the
projected total $n_{DOS}(\epsilon)$ of the Ni$_7$ cluster calculated
without taking into account the relativistic effects. As one can
see from the spectrum, there are the free electronic states higher
the LUMO level, which describe well the calculated spin DOS $S(\epsilon)$
of Ni$_7$ clusters (see the Fig.\ref{fig.3}).

\begin{figure}[tbp!]                                                                                                            \begin{center}                                                                                                                  \includegraphics[height=12cm, width=10cm]{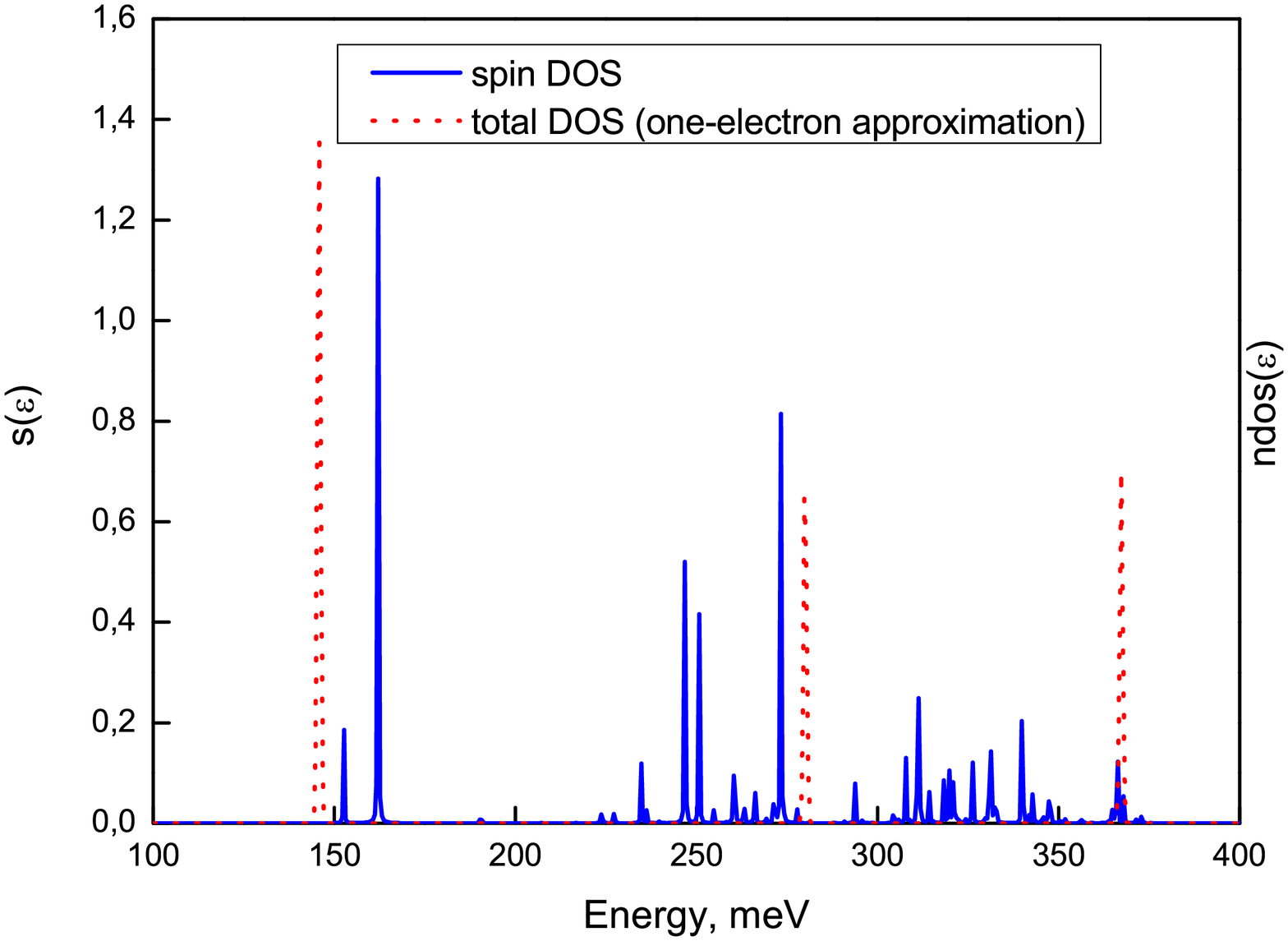}                                                                        \end{center}                                                                                                                    \caption{(Color online) The comparison of the projected spin DOS $S(\epsilon)$ of Ni$_7$ cluster on a silicon surface and       one-electron $n_{DOS}(\epsilon)$ of Ni$_7$ cluster.} \label{fig.4}                                                              \end{figure}

The comparison of the projected spin DOS $S(\epsilon)$ of Ni$_7$ cluster
on a silicon surface and the one-electron
$n_{DOS}(\epsilon)$ of Ni$_7$ cluster is shown on the Fig.\ref{fig.4}.
In this Figure the Ni 3d-hole one-electron cluster states which coincide with low-lying (0 - 500 mеV) excited spin states have been allocated.

The properties of the 3d-electrons are best probed in an X-ray absorption
experiment by the excitation of $2p$ core electrons to
unfilled $3d$ states. In principle, X-ray absorption spectra contain
contributions from both $p\to d$ and $p\to s$ transitions,
but in practice the $p\to d$ channel dominates by a factor $>$20.
The line intensities, denoted $I_{L_2}$ and $I_{L_3}$,
respectively, are directly proportional to the number of $d$ holes.
The use of circularly polarized X-rays opens the door for
spin studies (XMCD spectroscopy). In this methodology                                                                           \begin{equation}                                                                                                                {\bf M}_s=-2\langle \hat {\bf S}\rangle \mu_B/\hbar=(N_{\uparrow} - N_{\downarrow})\mu_B,                                       \end{equation}                                                                                                                  where $N_{\uparrow}$ and $N_{\downarrow}$ are the numbers of $d$ holes
with the spin $\uparrow$ and the spin $\downarrow$.
Now the two-spin model is the main model in XMCD \cite{XMCD}.
However it is surprising that the treatment of the X-ray absorption
spectra, in general, is based on the simplified model neglecting a mixing
of the spin states. Undoubtedly, it would be useful to
reconsider interpretation of these spectra, already
taking into account mixing of spin states.

In this work X-Ray Absorption Near-Edge Structure (XANES) spectroscopy was
used both to study the adsorption geometry and to get information concerning
the electronic properties of the deposited Ni-clusters.
An interpretation of the experimental XANES data, which were taken from \cite{XANES10},
was provided in the present study by model spectra, based on real-space
multiple-scattering approach implemented in the FEFF9 program code \cite{XANES3,XANES4}.
FEFF9 uses an ab initio self-consistent real space multiple scattering
approach, including polarization dependence, core-hole effects, and local
field corrections, based on self- consistent, spherical muffin-tin scattering potentials.

\begin{figure}[tbp!]                                                                                                            \begin{center}                                                                                                                  \includegraphics[height=12cm, width=10cm]{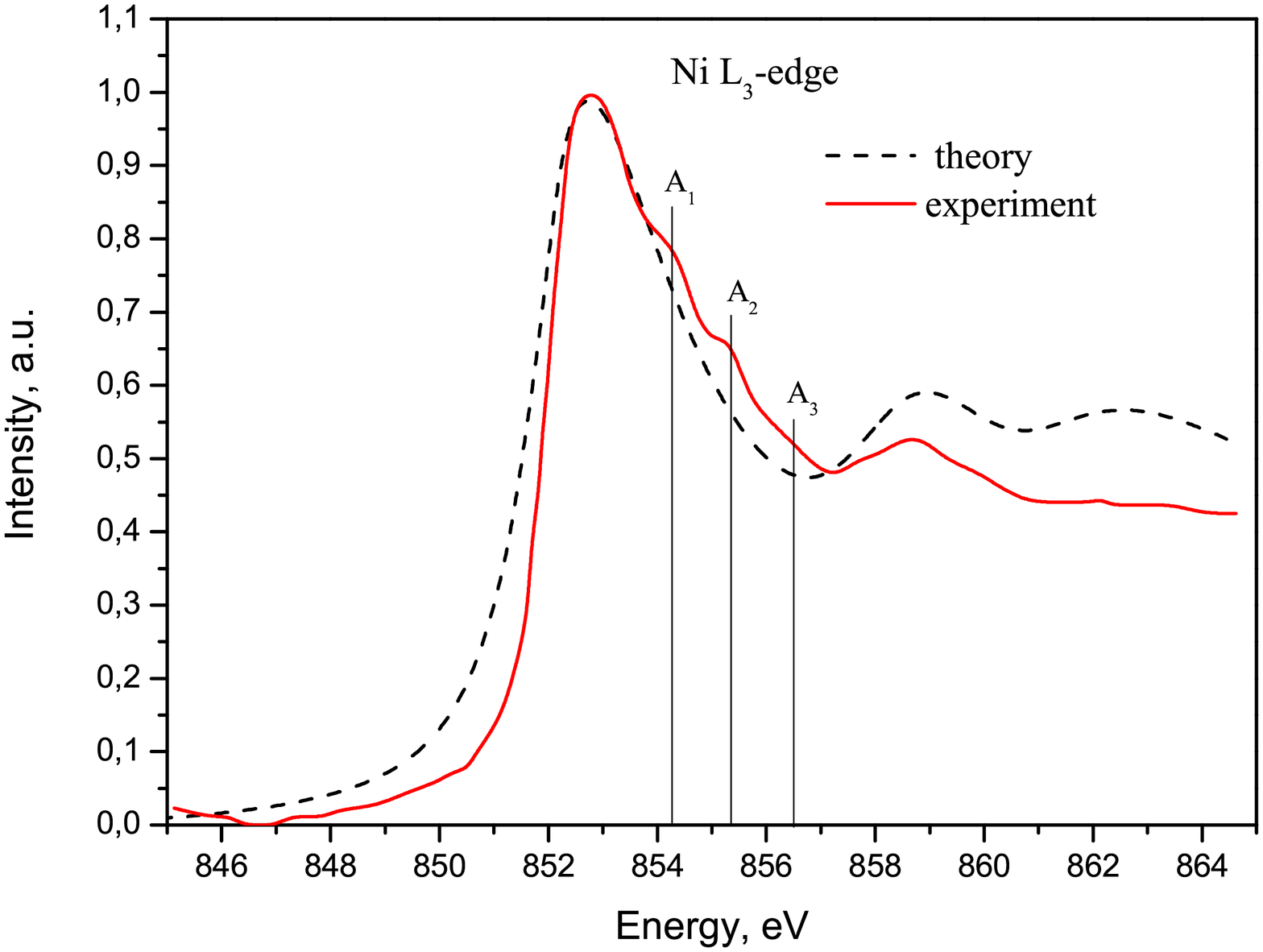}                                                                          \end{center}                                                                                                                    \caption{(Color online) The theoretical one-electron and experimental NiL$_3$-edge
XANES for Ni$_7$ cluster on a silicon surface.} \label{fig.8}                                                                                                         \end{figure}

\begin{figure}[tbp!]                                                                                                            \begin{center}                                                                                                                  \includegraphics[height=12cm, width=10cm]{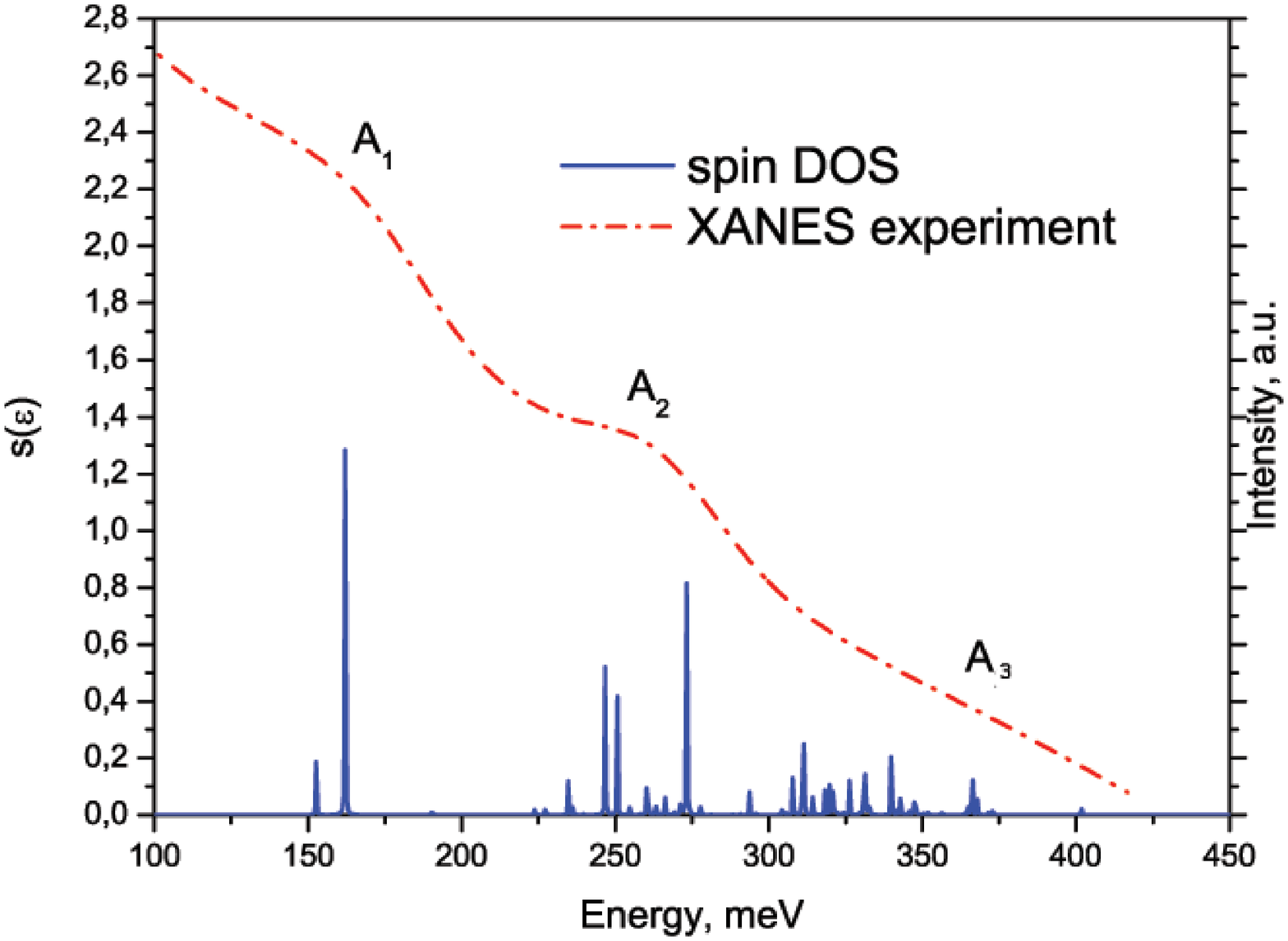}                                                                         \end{center}                                                                                                                    \caption{(Color online) The comparison of the calculated spin DOS $S(\epsilon)$ of the Ni$_7$
cluster on a silicon surface with the experimental Ni L$_3$-XANES spectra of carbon
nanotube with Ni catalyst nanoparticle taken from \cite{XANES2}. The experimental spectrum was shifted on the value of the Ni 2p$_{3/2}$ binding energy.} \label{fig.9}
\end{figure}

On the Fig.\ref{fig.8} we show the comparison between the theoretical spectrum,
calculated in the one-electron approximation, with the experimental one,
taken from \cite{XANES10}. The experimental spectrum was measured for the similar system of Ni
clusters grown on the carbon nanotube layer.
We have choose the experimental data as the most suitable for our system Ni clusters on
the silicon surface because silicon and carbon atoms have the similar configuration of the
electronic shell - [Ar]$ns^2np^3$, where n$=$2 and 3 for carbon and silicon, respectively.
As one can see, there are three features
on the experimental spectrum called $A_1$, $A_2$ and $A_3$, which represent the spin structure
of the Ni clusters. This structure can’t be repeated without taking into account of the
multiplet effects. This fact is obvious from the theoretical spectrum calculated by the
one-electron FEFF9 code.
If compare the spin DOS $S(\epsilon)$ of Ni$_7$ cluster on a silicon surface with the
post-edge region of the experimental Ni $ L_3 $-XANES spectrum for Ni catalyst
nanoparticles (see Fig.\ref{fig.9}) one can notice that the energy position of
the features of spin DOS $S(\epsilon)$ coincides with the post-edge features of
the absorption spectrum of Ni clusters.                                                                                                                               


\begin{figure}[tbp!]                                                                                                            \begin{center}                                                                                                                  \includegraphics[height=23cm, width=8.5cm]{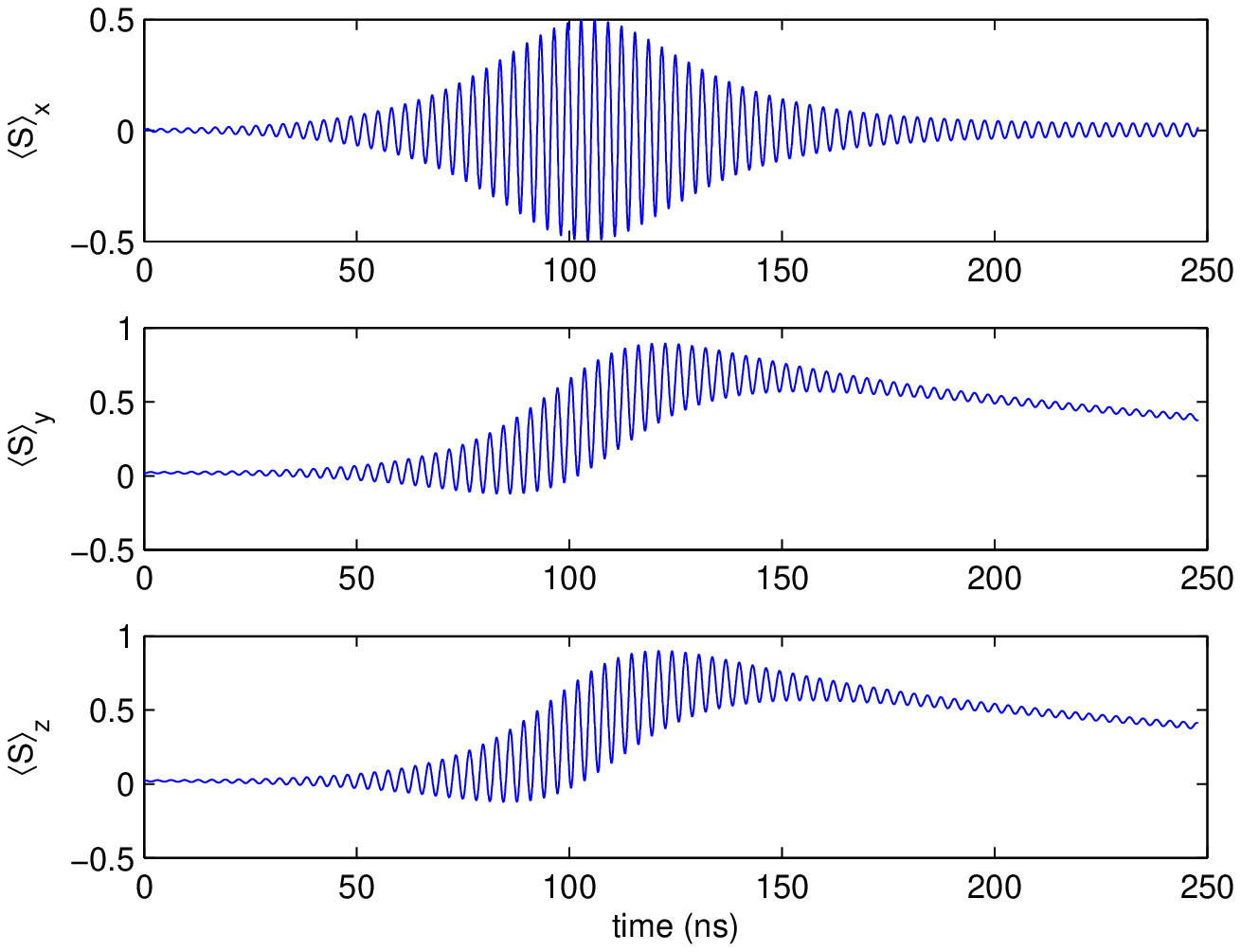}                                                                         \end{center}                                                                                                                    \caption{(Color online) Time evolution of the quantum mechanical expectation
$\langle S\rangle_{x,y,z}$ of Ni$_7$ cluster on a  silicon surface ($H_z$=0.01$Ts$).} \label{fig.10}                                                                               \end{figure}
\begin{figure}[tbp!]                                                                                                            \begin{center}                                                                                                                  \includegraphics[height=23cm, width=8.5cm]{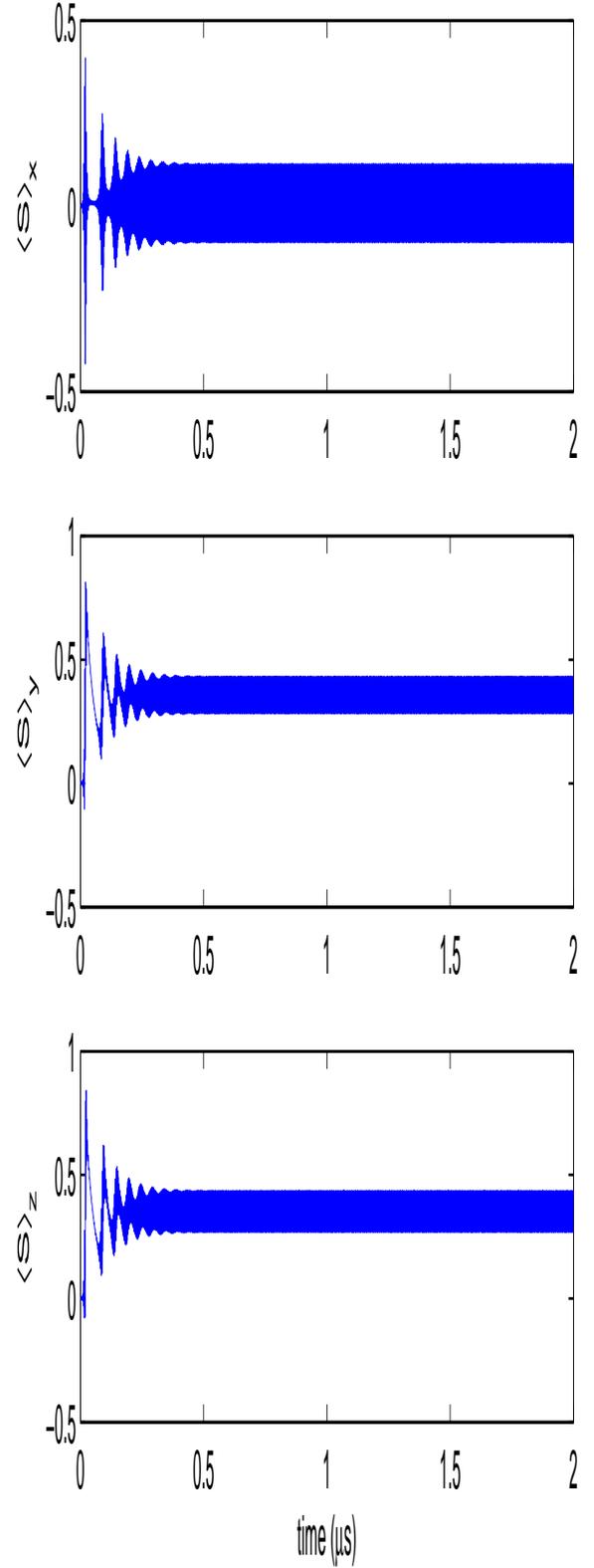}                                                                        \end{center}                                                                                                                    \caption{(Color online) Time evolution of the quantum mechanical expectation
$\langle S\rangle_{x,y,z}$ of Ni$_7$ cluster on a  silicon surface ($H_z$=1.76$Ts$).} \label{fig.11}                                                                               \end{figure}

A studying of the decoherence in the solid-state spin-based qubit systems
was the focus of our project. Since the
decoherence is a complex many-body non-equilibrium process,
and its description by purely analytical
means is rarely possible, our main tool was the direct and
highly accurate numerical solution of the time-dependent
Schr\"odinger equation for the whole system
(qubits plus their silicon substrate). This is a very
difficult but extremely reliable approach, involving no
approximations about a system or a environment.
In this study we have developed and used the advanced
numerical techniques to accurately assess the details of the
decocerence process governing the dynamics of SSNQ interacting
with a silicon surface.
A well-studied model for the decoherence is a central-spin
system coupled to the N noninteracting spins. The exact
evaluation of dynamics for the central spins is obtained for
the special cases, where there is uniform coupling with the
 spin bath, or a special choice of the initial states, or the
system-bath interaction is a simple interaction
between the $z$-component of the spins. Each such studied case can explain the experimentally observed results in some solid-state devices. We offer a model in which there
is no division of the Hamiltonian on $H_0$ and $H_{bath}$, and
the interaction from a bath (environment) is considered in the
GSH technique at the calculation of the exchange integrals $J_{ij}$
of a nanosystem Ni$_7$-Si in the one-electron approach taking
into account the chemical bonds of all atoms of the Si substrate
(environment) with the atoms of the Ni$_7$ cluster. We investigated in
detail the electron spin decoherence
and Rabi oscillations for the various magnetic fields $H_z$.
As a result, we developed a new method of simulations of the decoherence, which allowed
modeling of the realistically large systems (thousands of a quantum spins)
with a complex dynamics. Here we used 2187 spins for the
generalized spin Hamiltonian. A noticeable part of our studies
has become possible due to the progress in the methods for a numerical
modeling of a decoherence. We also studied the visibility and
the decay of Rabi oscillations in SSNQ qubits. In Fig.\ref{fig.10} we
have plotted the Rabi oscillations to evaluate the qubit as a function of
the time and $H_z$=0.01$Ts$. An important requirement for
a quantum computing is a control, which can be made through the magnetic fields.
In Fig.\ref{fig.11} we have plotted the Rabi
oscillations to evaluate the qubit as a function of the time and $H_z$=1.76$Ts$.
Here we shows the quantum state stabilization
(stabilizing Rabi oscillations) in the Ni$_7$--Si nanosystem for $H_z$=1.76$Ts$. Thus, we have observed the stabilized Rabi oscillations and the stabilized quantum dynamical qubit state and the Rabi driving after the fixed time (0.327 $\mu s$).

Today, in fact, the quantum entanglement is recognized as a new physical resource
which is important for a quantum computation. We have analyzed
a behavior of the entanglement in the finite clusters
of the quantum spins and have shown that the entanglement in these
systems is significantly modified near special values of the
energy and temperature. These results could be particularly relevant for applications in quantum computations. In this paper we have identified the global entanglement patterns in SSNQ by analyzing  the behavior of the density of entanglement as a function of the energy and temperature, which ultimately leads to a change in the spin structure of the cluster.
\begin{figure}[tbp!]                                                                                                           \begin{center}                                                                                                                  \includegraphics[height=9cm, width=9.5cm]{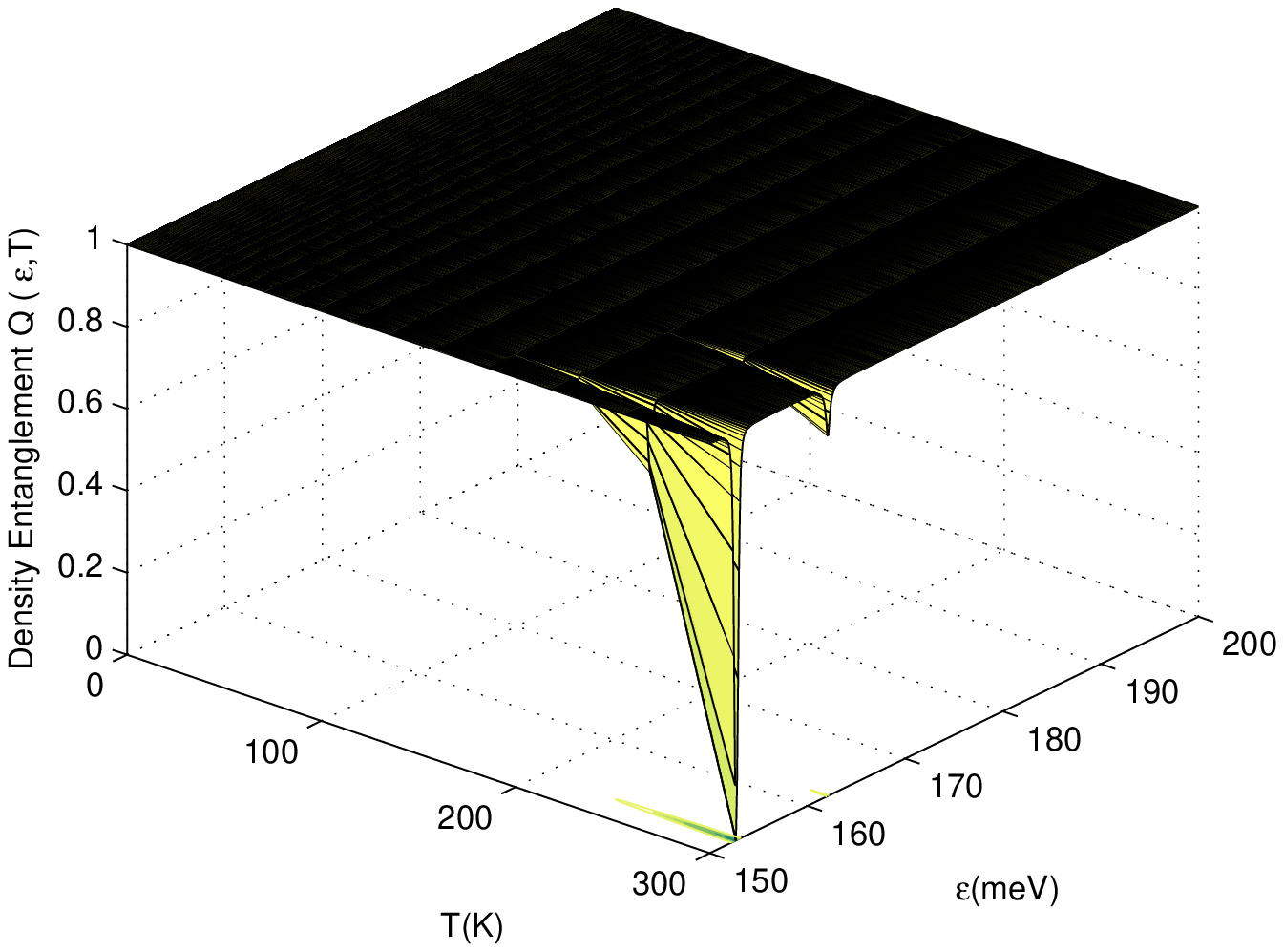}                                                                 \end{center}                                                                                                                    \caption{(Color online) The calculated density of global entanglement                                                           vs temperature and energy for Ni$_7$ cluster on a silicon surface.} \label{fig.12}                                              \end{figure}
Identifying and measuring entanglement in the multi-spin systems is
a challenge. Apart from the case of a two-qubit
system, where an entanglement can be identified both for a pure and
a mixed state \cite{Woo28,Uhl29,Aud30,Woo31},
the multi-qubit entanglement is an open problem and to date several
measures of an entanglement have been proposed \cite{VC32,WN33,CH34}.
For the analysis that follows we will be using the global entanglement
\cite{CH34} since this will enable us to deal with
the many spin eigenvalues (spin modes). Since we will assume only a single excitation
in each cluster spin mode, we can treat the cluster spin
states as a set of qubits for the purpose of the computing global entanglement.
For quantifying the distribution of an entanglement
between the individual spin eigenvalues in the spin structure of the SSNQ-system
in a Ni$_7$-cluster on a silicon substrate we use the
density of entanglement. In Fig.\ref{fig.12} we have plotted
the calculated density of the global entanglement vs temperature and
energy for the Ni$_7$ cluster on a silicon surface. From the Figure we can see
that in the wide interval of the temperatures from 0 to
230$K$ and in the range of the energy from 0 to 200 meV practically all
states of SSNQ  have the most entanglement that is very important
for the work of the quantum computer. Since 230$K$ at the energy
$\epsilon_1=153$ meV and $\epsilon_2=162$ meV there is a sharp
falling of a complexity and we observed the spectral gap in DOE.
In this context, the systematic studies of the relationship between
the amount and nature of an entanglement and a spin strucrure has
been pursued in order to the identify optimal researches to the create specific types of the entanglements.
\section{CONCLUDING REMARKS}
Under technological inputs, the cluster magnetism is now moving more
and more towards surface science with
the implications for the use of new theoretical and experimental
techniques and with the development of new synthetic approaches.
The cluster nanomagnets on a nonmagnet substrate have specific features
that make them paradigmatic cases to the test models and with
which we may build the novel quantum architectures. Carrying out a theoretical
and experimental investigation of the quantum model
of the solid-state spin qubit Ni on a nonmagnetic silicon surface for
the quantum register--the quantum computer theory is developed. Within
the present article, we used the numerical simulations and X-ray spectral methodology
to study of a spin structure, micromagnetic
simulations decoherence and a global density of entanglement
in the N-spin qubit systems. We have also investigated important
fundamental problems of an entanglement and a decoherence theory
in the N-spin nanosystem Ni3d--Si heterostructure. We have studied the
spin structure, the dynamics of decoherence vs magnetic field and the global
entanglement in the hybrid nanosystems. As a result, we have
provided the detailed theoretical and experimental description
for many important aspects of the spin structure, the decoherence and
the entanglement in the systems relevant for quantum information processing
and quantum computers. No doubt, the challenges we face in
building a real siliconbased N-qubit with N$\geq$2000 are
significant, but our initial results offer hope that
large-scale quantum computing may one day be realized.                                                                                                                                                  
\section{ACKNOWLEDGMENTS}
The part of this research (for V.M.) has been supported by Russian Ministry of Science, Grant No. 213.01-11/2014-6.


\begin{thebibliography}{99}
%
\bibitem{i1}
M. A.Nielsen, I. L.Chuang, Quantum computation and quantum information, Cambridge
University Press, Cambridge, New York, 2000.
%
\bibitem{i2} D.Loss and D. P.DiVincenzo, Phys. Rev. A{\bf 57},120 (1998).
%
\bibitem{i21} G.Burkard, D.Loss, and D. P.DiVincenzo, Phys. Rev. B{\bf 59},2070 (1999).
%
\bibitem{i3} B. E.Kane, Nature {\bf 393},133 (1998).
%
\bibitem{i4} J.Wrachtrup and F.Jelezko, J.Phys.: Cond. Matter {\bf 18},807 (2006).
%
\bibitem{i41} L.Childress, M. V.Gurudev, J. M.Taylor, Science {\bf 314},281 (2006).
%
\bibitem{i5} P.Shor, Proceedings of the 35th Annual Symposium on Foundations of Comduter Science,
             1994, pp. 116-123.
%
\bibitem{i6} K.Wu, B.Zhou and W.Cao, Phys. Lett.A{\bf 362},381 (2007).
%
\bibitem{i7} Gary Wolfowicz, Alexei M.Tyryshkin, Richard E.George, Helge Riemann, Nikolai V.Abrosimov,
             Peter Becker, Hans-Joachim Pohl, Mike L.W.Thewalt, Stephen A.Lyon and John J.L.Morton,
             Nat. Nanotech.{\bf 8},561 (2013).
%
\bibitem{i8} Filippo Troiania and Marco Affronte, Chem.Soc.Rev.{\bf 40},3119 (2011).
%
\bibitem{i9} Oleg O.Brovko, Oleg V.Farberovich and Valeri S.Stepanyuk, J.Phys.:Condens. Matter{\bf 26},315010 (2014).
%
\bibitem{i10} Volodymyr V.Maslyuk, Ingrid Mertig, Oleg V.Farberovich, Alex Tarantul, and Boris Tsukerblat,
              Eur.J.Inorg.Chem.{\bf 2013},1897 (2013).
%
\bibitem{i11} W.Zhang, N.Konstantinidis, K. A.Al-Hassanieh, and V. V.Dobrovitski,
              J.Phys.:Cond.Matter{\bf 19},083202 (2007).
%
\bibitem{i12} K. A.Al-Hassanieh, V. V.Dobrovitski, E.Dagotto, and B. N.Harmon,
              Phys. Rev. Lett.{\bf 97},037204 (2006).
%
\bibitem{i13} W.Zhang, K. A.Al-Hassanieh, V. V.Dobrovitski, E.Dagotto, B. N.Harmon,
              Phys. Rev.B{\bf 74},205313 (2006).
%
\bibitem{i14} R.Horodecki, P.Horodecki, M.Horodecki and K.Horodecki,
              Rev.Mod.Phys.{\bf 81},865 (2009).
%
\bibitem{i15} L.Amico, R.Fazio, A.Osterloch and V.Vedral, Rev.Mod.Phys.{\bf 80},517 (2008).
%
\bibitem{i16} G. L.Kamta and A. F.Starace, Phys.Rev.Lett.{\bf 88},107901 (2002).
%
\bibitem{i17} C. H.Bennett and D. P.DiVincenzo, Nature {\bf 404},247 (2012).
%
\bibitem{GSH19} J.J.Borras-Almenar, J.M.Clemente-Juan, E.Coronado, and B.S.Tsukerblat,
                Inorg.Chem.{\bf 38},6081 (1999).
%
\bibitem{i18} Xiao-Zhong Yuan, Hsi-Sheng Goan, and Ka-Di Zhu, New J.Phys.{\bf 13},023018 (2011).
%
\bibitem{ITO20} J.J.Borras-Almenar, J.M.Clemente-Juan, E.Coronado and B.S.Tsukerblat,
                J.Comput.Chem.{\bf 22},985 (2001).
%
\bibitem{i161} G.Lagmago Kamta, A. Y.Istomin, and A. F.Starace, Eur. Phys. J. D{\bf 44},389 (2007).
%
\bibitem{XANES1} V.L.Mazalova,A.A.Guda,N.Yu.Smolentsev,O.E.Polozhentsev,
                 I.G.Alperovich,A.V.Soldatov,S.P.Lau,X.H.Ji,In:Piezoelectrics and Related Materials:
                 Investigations and Applications, Ed.I.A.Parinov,Nova Science Pub Inc., p.51-77,2012.

\bibitem{XANES2} Victoria Mazalova,Alexander Soldatov,Roy Johnston,
                 Alexei Yakovlev,Thomas Moller,Sorin Adam, J.Phys.Chem. C{\bf 113},9086 (2009).
\bibitem{XANES3} J.J.Rehr, J.J.Kas, F.D.Vila, M.P.Prange, K.Jorissen, J.Phys.Chem. {\bf 12},5503 (2010).
%
\bibitem{XANES4} J.J.Rehr, J.J.Kas, M.P.Prange, A.P.Sorini, Y.Takimoto, F.D.Vila,
                 Comptes Rendus Physique {\bf 10},548 (2009).
\bibitem{XANES5} G. te Velde, F.M.Bickelhaupt, S.J.A. van Gisbergen, C.Fonseca Guerra,
                 E.J.Baerends, J.G.Snijders and T.Ziegler,J.Comp.Chem. {\bf 22},931 (2001).
\bibitem{XANES6} C.Fonseca Guerra, J.G.Snijders, G. te Velde and E.J.Baerends,
                 Theor.Chem.Accoun.{\bf 99},391 (1998).
\bibitem{XANES7} A.D.Becke, Phys.Rev. A{\bf 38},3098 (1988).
\bibitem{XANES8} C.Lee, W.Yang and R.G.Parr, Phys.Rev. B{\bf 37},785 (1988).
\bibitem{XANES10} J.Gao, J.Zhong, L.Bai, J.Liu, G.Zhao and X.Run, Scientific Report {\bf 4}, 3606 (2013).
\bibitem{MasFar21} D. W.Boukhvalov, V. V.Dobrovitski, M. I.Katsnelson, A. I.Lichtenstein, B. N.Harmon, P.K\"ogerler,
                   Phys. Rev. B{\bf 70},054417 (2004).

\bibitem{Kub22} J.K\"ubler, K.-H.H\"ock, J.Sticht, A. R.Williams,
                J. Phys. F{\bf 18},469 (1988).

\bibitem{Mas23} V.V.Maslyuk, Ph. D. Thesis, Martin Luther University of Halle-Wittenberg, 2009.

\bibitem{LDiV24} D.Loss and D.P.DiVincenzo,
                 Phys. Rev. A{\bf 57},120 (1998).

\bibitem{W25} R.Wieser, Phys.Rev. B{\bf 84},054411 (2011).

\bibitem{W26} R.Wieser, Phys.Rev.Lett. {\bf 110},147201 (2013).

\bibitem{W27} R.Wieser, E. Y. Vedmedenko, and R. Wiesendanger,
              Phys.Rev.Lett. {\bf 106}, 067204 (2011).

\bibitem{Woo28} W. K.Wootters, Phys. Rev. Lett. {\bf 80},2245 (1998).

\bibitem{Uhl29} A.Uhlmann, Phys.Rev. A{\bf 62},032307 (2000).

\bibitem{Aud30} K.Audenaert, F.Verstraete and De Moor,
                Phys. Rev. A{\bf 64},052304 (2001).

\bibitem{Woo31} W. K.Wootters, Quantum Inf. Comp. {\bf 1},27 (2001).

\bibitem{VC32} Valerie Coffman, Joydip Kundu, and William K. Wootters,
               Phys.Rev. A{\bf 61},052306 (2000).

\bibitem{WN33} Alexander Wong and Nelson Christensen,
               Phys.Rev. A{\bf 63},044301 (2001).

\bibitem{CH34} Chang-Shui Yu and He-shan Song,
               Phys.Rev. A{\bf 71},044301 (2005).

\bibitem{MW35} D.A.Meyer and N.R.Wallach,
               J.Math.Phys. {\bf 43},4273(2002).

\bibitem{Bren36} G.K.Brennen,
                 Quantum.Inf.Comp. {\bf 3},619 (2003).

\bibitem{LG37} C.Lazarou, B.M.Garraway, J.Piilo, S.Maniscalco,
               J.Phys.B:At.,Mol.and Opt.Phys. {\bf 44},6 (2011).

\bibitem{TM38} N.Troullier, J.L.Martins,
               Phys.Rev. B{\bf 1991},43 (1993).

\bibitem{Per39} J. P. Perdew, K. Burke, M. Ernzerhof,
                Phys. Rev. Lett. {\bf 77},3865 (1996).

\bibitem{SIESTA} D.Sanchez-Portal, P.Ordejon, E.Artacho, J.M.Soler,
                 Int.J.Quantum Chem. {\bf 65},453 (1997).

\bibitem{Kort18} J.Kortus, M.R.Pederson, T.Baruah, N.Bernstein, C.S.Hellberg, Polyhedron {\bf 22},1871 (2003).

\bibitem{Bor42} S.Bornemann, O.Sipr, S.Mankovsky, S. Polesya, J. B. Staunton, W. Wurth, H. Ebert, and J. Minar,
                Phys.Rev. B{\bf 86},104436 (2012).

\bibitem{XMCD} J.St\"ohr,
               J.Mag.Mag.Mat.{\bf 200},470 (1999).

\end{thebibliography}
\end{document}